\documentclass[sigconf,authorversion]{acmart}
\makeatletter
\@ifundefined{iflatexml}{\newif\iflatexml\latexmlfalse}{}
\makeatother
\usepackage{multirow}
\usepackage{array} 
\usepackage{arydshln} 
\usepackage{bm}
\usepackage{mathtools}
\usepackage{url}

\AtBeginDocument{%
  }

\copyrightyear{2026}
\acmYear{2026}
\setcopyright{cc}
\setcctype{by}
\iflatexml
\else
\acmConference[ICMI '26]{INTERNATIONAL CONFERENCE ON MULTIMODAL INTERACTION}{October 05--09, 2026}{Napoli, Italy}
\acmBooktitle{INTERNATIONAL CONFERENCE ON MULTIMODAL INTERACTION (ICMI '26), October 05--09, 2026, Napoli, Italy}
\acmDOI{10.1145/3776574.3831145}
\acmISBN{979-8-4007-2318-6/2026/10}
\fi

\begin{document}

\title{Real-time Generation of Listener Nodding\\via Prediction of Kinematic Parameters\\for Avatar Dialogue Systems}

\author{Kazushi Kato}
\affiliation{%
  \institution{Kyoto University}
  \country{Japan}}
\email{katou@sap.ist.i.kyoto-u.ac.jp}

\author{Koji Inoue}
\affiliation{%
  \institution{Kyoto University}
  \country{Japan}}
\email{inoue@sap.ist.i.kyoto-u.ac.jp}

\author{Taiga Mori}
\affiliation{%
  \institution{Kyoto University}
  \country{Japan}}
\email{mori@sap.ist.i.kyoto-u.ac.jp}

\author{Divesh Lala}
\affiliation{%
  \institution{Kyoto University}
  \country{Japan}}
\email{lala@sap.ist.i.kyoto-u.ac.jp}

\author{Tatsuya Kawahara}
\affiliation{%
  \institution{Kyoto University}
  \country{Japan}}
\email{kawahara@i.kyoto-u.ac.jp}

\renewcommand{\shortauthors}{Kato et al.}

\begin{abstract}
In human dialogue, we achieve smooth communication by expressing nonverbal cues such as eye contact, nodding, and facial expressions with precise timing. It is expected for conversational avatars to express these cues appropriately to realize natural and human-like interactions. This study focuses on nodding, which is crucial for demonstrating active listening and encouraging further user utterances. We propose a model that predicts both timing and kinematic parameters representing the motion features of listener nodding in real time. The proposed model consists of a timing prediction module and a kinematic parameter prediction module. Each implements a dyadic attention network over the speaker and listener channels based on the technique of Voice Activity Projection (VAP). Unlike conventional models, this approach enables real-time prediction of kinematic parameters based on the specific context of the dialogue rather than just predicting the timing. Furthermore, we demonstrate the effectiveness of fine-tuning the kinematic parameter prediction module initialized from the trained timing prediction module. The proposed model is lightweight and capable of real-time operation, and it has been integrated into an avatar dialogue system. Subjective evaluation experiments shows that our proposed method significantly outperforms both a baseline with stochastic timing and another with fixed-motion nodding. The code and trained models are available at \url{https://github.com/MaAI-Kyoto/MaAI}.
\end{abstract}

\iflatexml
\else

\fi



\maketitle

\iflatexml
\noindent\textit{Accepted to ICMI '26}, Napoli, Italy, October 05--09, 2026.
DOI: \url{https://doi.org/10.1145/3776574.3831145}
\fi

\section{Introduction}

In human dialogue, we perceive nonverbal cues such as eye contact, nodding, and facial expressions from our interlocutors and express them for smooth communication.
It is expected for spoken dialogue systems and conversational avatars to express these cues appropriately to realize natural and human-like interactions.
In particular, while the user is speaking, timely nonverbal listener responses such as verbal backchannels and nodding should show that the system is listening and encourage the user to continue speaking.

Recently, research on generating listening head motions for conversational avatars has been conducted~\cite{difflistener2025, listenformer2024, siniukov2025ditailistener}, including causal real-time models~\cite{mikawa2024,zhao2025realistener} and integrations into dialogue systems~\cite{mikawa2024}.
Some approaches generate motions according to the avatar's role, such as listener or speaker, by modeling turn-taking and conversational state in the architecture~\cite{guo2025arig, ki2026avatarforcing}.
However, many of these models do not explicitly model the timing at which the listener can respond (i.e., backchannel opportunity points (BOPs)~\cite{gratch2006virtualrapport} or backchannel relevance places~\cite{noguchi2001}), or model it only indirectly, posing challenges for generating appropriately timed nodding backchannels.

\begin{figure*}[t]
  \centering
  \includegraphics[width=\linewidth]{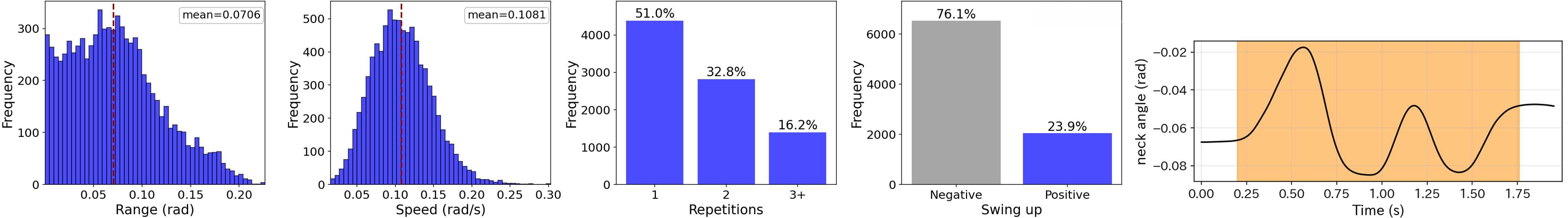}
  \Description{Left: histograms and bar charts of range, speed, repetitions, and swing-up for nodding events. Right: line plot of vertical neck angle over time with a shaded detected nodding segment.}
  \caption{Distributions of kinematic parameters (\emph{left}) and smoothed motion data with a detected nodding segment (\emph{right}).}\label{fg:param_dist}\label{fg:nod_detect}
\end{figure*}

Nodding is a nonverbal response in which participants shake their heads vertically during dialogue, primarily performed by the listener~\cite{maynard1987headmovement}.
It is often performed simultaneously with verbal backchannels such as ``um'' or ``uh-huh'', as a backchannel, and is closely related to turn-taking~\cite{mori2020, maynard1987headmovement, otsuka2020headmultifunction}.
Moreover, listeners use various forms of nodding depending on the dialogue context, thereby facilitating smooth communication~\cite{blomsma2024backchannelidiosyncratic, mori2021, mori2022cognitivenods}.
% For instance, nodding that accompanies a cognitive shift such as reception of new information or surprise tends to involve swing-up, and nodding expressing acceptance or acknowledgment tends to have a small range of movement, whereas nodding expressing surprise, interest, or understanding tends to have a large range of movement~\cite{mori2020, mori2021}.
Several models have been proposed for predicting nodding in dialogue systems, including models based on sequential probabilistic models~\cite{morency2010probabilisticbackchannel}, models using human-human dialogue data as surrogate data~\cite{noguchi2024attentivelisteningrobot}, and models that predict three types of nodding continuously in real time~\cite{kato2025realtimenodding}.
However, most of these models are limited to binary prediction of nodding versus not nodding, or prediction of predefined nodding classes, and systems integrating them can only replay fixed-pattern nodding rather than expressing diverse forms~\cite{jain2021exploringsemisupervised}.

In this paper, we propose a model that predicts in real time both nod timing and kinematic parameters of listener nodding, such as the number of repetitions and motion range (Figure~\ref{fg:param_dist}).
To model appropriate listener nod timing, the proposed model learns the onset timing of nodding events from annotated dialogue data.
In addition, by learning kinematic parameters for each ground-truth interval, the model can predict the form of nodding depending on dialogue context.
To the best of our knowledge, no model has been proposed that continuously predicts both timing and kinematic parameters of nodding in real time.
The diversity of nodding forms is known to be closely related to the listener's cognitive state (e.g., understanding, agreement, notice, and surprise) and turn-taking~\cite{mori2021, mori2022cognitivenods}.
Therefore, by predicting not only the timing of nodding but also its kinematic parameters and expressing context-appropriate forms of nodding in dialogue systems, it is expected to produce human-like listener responses.

The proposed model consists of a timing prediction module and a parameter prediction module, each implementing a dyadic attention network over the speaker and listener channels based on Voice Activity Projection (VAP)~\cite{ekstedt2022}, which models turn-taking in dyadic dialogue.
Similar VAP-style dyadic architectures have been applied to real-time verbal backchannel prediction and real-time nodding prediction in prior work~\cite{inoue2025yeahunohcontinuous,kato2025realtimenodding}.
Building on these prior studies, we present a model with the following features.
First, it takes the speech signals of both the speaker and the listener as input to predict both the timing and kinematic parameters of listener nodding in an end-to-end manner.
Second, the timing prediction module is trained with multi-task learning on nodding prediction, turn-taking, and verbal backchannel prediction and detection tasks.
We then fine-tune the kinematic parameter module, initialized from the timing-trained module, leveraging turn-taking for appropriate nodding form.
Third, instead of classifying nodding into discrete categories and replaying a matched motion, the proposed model directly predicts kinematic parameters and synthesizes diverse and human-like head motion from them.
Finally, the proposed model is lightweight and capable of real-time operation, so integrating it into spoken dialogue systems or conversational avatars enables real-time generation of listener nodding.

\section{Dataset} \label{sec:data}

We use the attentive listening dataset collected through Wizard-of-Oz experiments with the android ERICA~\cite{kawahara2019}.
A trained actor conducted Japanese attentive-listening dialogues with elderly participants (everyday small talk) and university students (COVID-19-related difficulties).
Because the original recordings lacked diverse nodding, we annotated nodding intervals on additionally recorded listener gestures from a prior study~\cite{kato2025realtimenodding} and used them for training.
The dataset contains 90 dialogues averaging 8 minutes each (about 12 hours in total), and the transcripts include segment-level annotations of listener verbal backchannels.

\subsection{Nodding Annotation} \label{subsec:nod_annotation}

\begin{table}[t]
  \centering
  \setlength{\tabcolsep}{0.35cm}
  \caption{Distribution of nodding segments}\label{tb:nod}
  \begin{tabular}{lccc} \hline
    &Time (s)&Ratio (\%)&Count \\ \hline
    Nodding&15477.9&37.0&8581 \\
    No nodding&26350.9&63.0& \\ \hline
  \end{tabular}
\end{table}

We detected nodding from the vertical neck angle (radians) in the motion data.
The motion data were resampled to 120~Hz, smoothed using a 15-frame moving average filter and a 4th-order Butterworth low-pass filter, and nodding segments were detected using gradient-based rules.
We manually corrected the detected segments by reviewing the motion data (Figure~\ref{fg:nod_detect}, right).
Nodding segments account for 37.0\% of the total time (Table~\ref{tb:nod}).
The large ratio of nodding is attributed to the nature of attentive-listening dialogues.

\subsection{Kinematic Parameters} \label{subsec:kinematic_params}

Based on previous studies that analyzed the motion features of nodding~\cite{mori2021, mori2022cognitivenods, mori2025structure}, we define the following four kinematic parameters and extract them from each detected nodding event.

\begin{itemize}
    \setlength{\itemsep}{0pt}
    \setlength{\parskip}{0pt}
    \item \textbf{Range} (rad): Represents the magnitude of the head movement during a nod. Computed as the absolute difference between the head angle at the segment onset and the angle at the lowest point.
    \item \textbf{Speed} (rad/s): Represents the velocity of head movement during a nod. Computed as the mean absolute angular velocity over the entire nodding segment.
    \item \textbf{Repetitions}: Represents the number of nodding cycles. A single cycle is defined as one consecutive pair of downward and upward head movements~\cite{mori2025structure}. Manually annotated for each nodding event while reviewing the motion data.
    \item \textbf{Swing-up}: Represents whether the nod begins with an upward head movement. Labeled when its motion starts with upward movement.
\end{itemize}

Figure~\ref{fg:param_dist} (\emph{left}) shows the distributions: mean range 0.0706~rad, mean speed 0.1081~rad/s, single-cycle nods 51.0\%, two-cycle nods 32.8\%, nods with three or more cycles 16.2\%, and swing-up at the start of the motion in 23.9\% of events (76.1\% without).

\section{Proposed Model} \label{sec:proposed}

This section describes the proposed model for predicting both timing and kinematic parameters of listener nodding in a continuous and real-time manner.
Following VAP-style multi-task learning for backchannel and nodding prediction~\cite{inoue2025yeahunohcontinuous,kato2025realtimenodding} and end-of-utterance prediction~\cite{ishii2025predictingeot}, our main extension is a dedicated module that directly predicts nodding kinematic parameters beyond timing or fixed classes.
We first describe the model architecture, then explain multi-task training for timing prediction, followed by fine-tuning for kinematic parameter prediction.

\subsection{Architecture} \label{sec:architecture}

\begin{figure}[t]
  \centering
  \includegraphics[width=\linewidth]{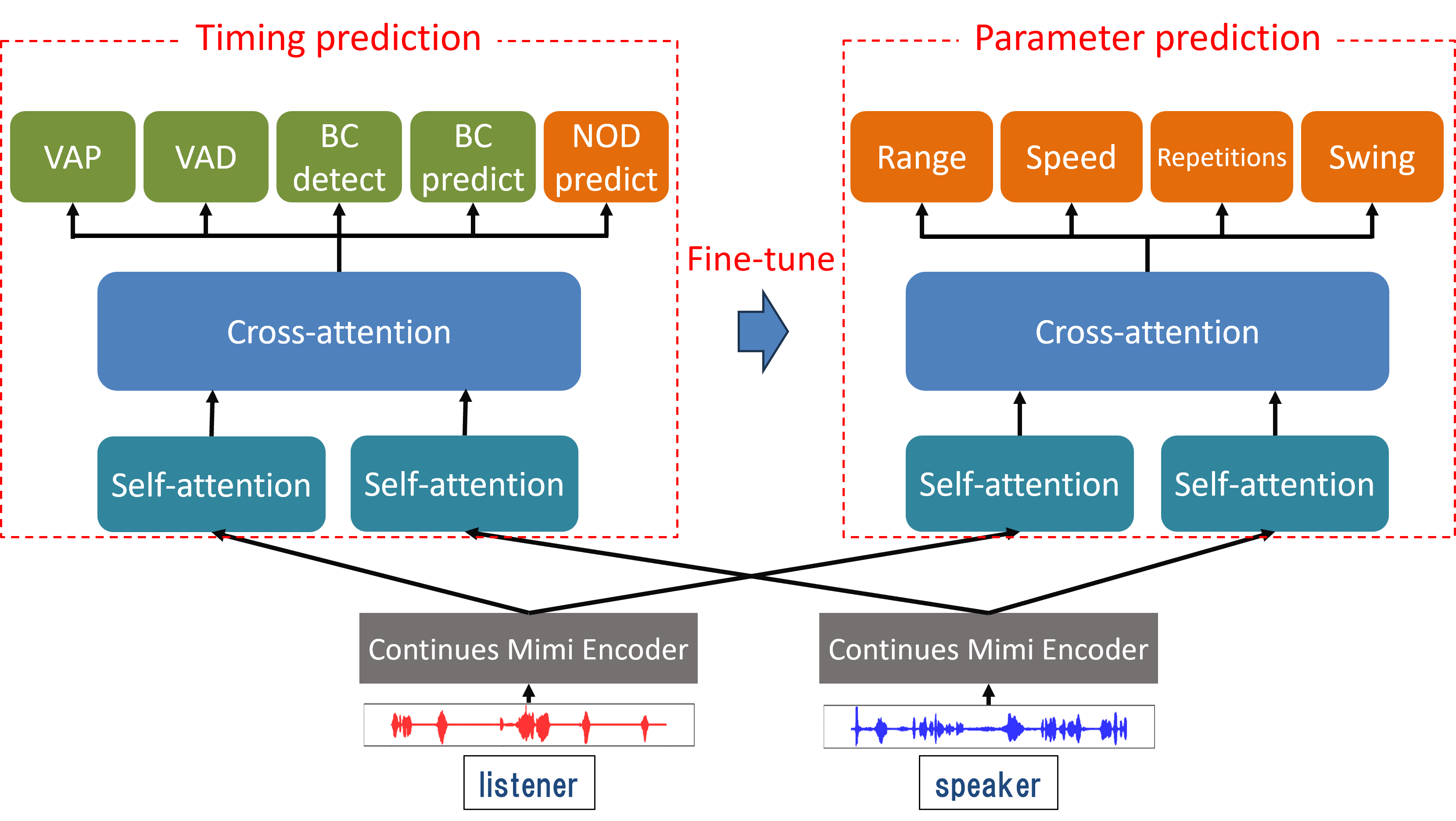}
  \Description{Block diagram of the proposed model: timing and parameter prediction modules with Continuous Mimi encoders and cross-attention.}
  \caption{Architecture of the proposed model}\label{fg:architecture}
\end{figure}

Figure~\ref{fg:architecture} shows the architecture of our proposed model consisting of a timing prediction module and a kinematic parameter prediction module.
The timing prediction task and the kinematic parameter prediction task differ in nature, so we allocate them to separate modules and fine-tune the parameter prediction module from the trained timing module.
Each module implements a dyadic attention network over the speaker and listener channels, following VAP~\cite{ekstedt2022}, which models turn-taking in dyadic dialogue.

We employ a component of the Mimi audio codec as an audio encoder, introduced in Moshi~\cite{defossez2024moshi}.
Specifically, we utilize the continuous features output by Mimi's neural encoder rather than the codebook-quantized discrete tokens.
This preserves causality while allowing predictions to rely on both acoustic and semantic information in the speech signal.
We refer to this as the Continuous Mimi Encoder.

In the proposed model, the audio waveforms of the two dialogue participants are each encoded using the Continuous Mimi Encoder, and the output features for each channel are processed by Self-attention Transformers in the respective modules.
These representations are then passed through Cross-attention Transformers, enabling reference to each other's attention states, and finally task-specific linear layers produce the predictions.

\begin{figure}[t]
  \centering
  \includegraphics[width=\linewidth]{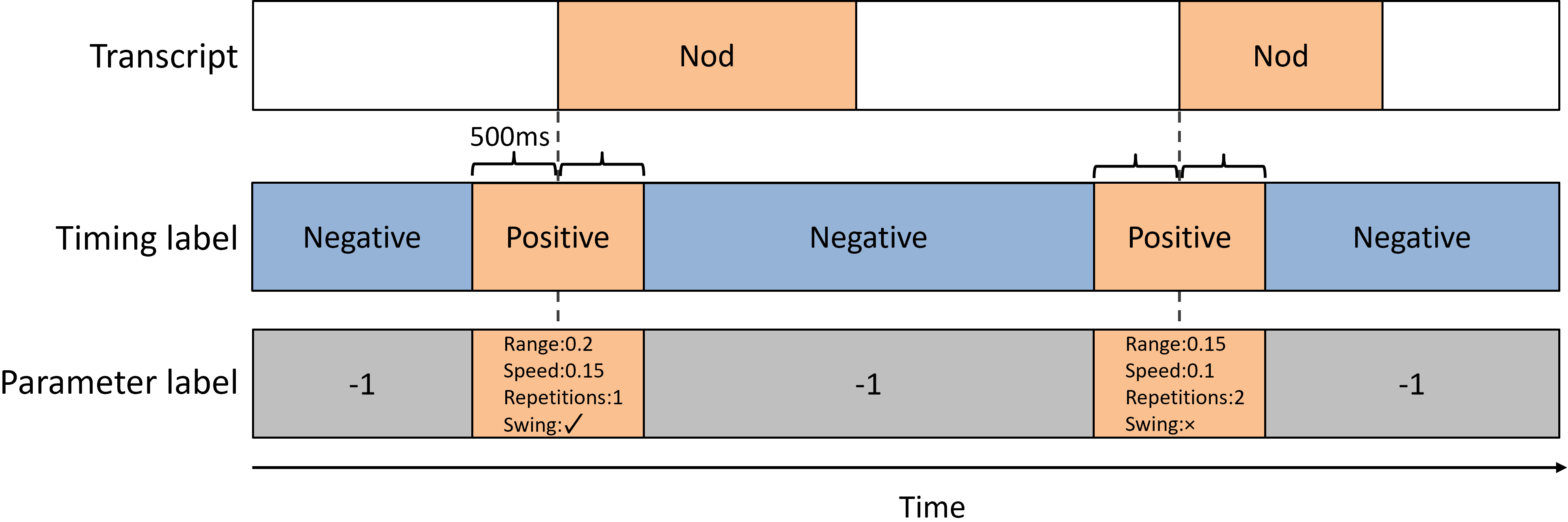}
  \Description{Illustration of ground-truth nodding label intervals relative to the nodding segment onset and transcript.}
  \caption{Definition of ground-truth labels}\label{fg:label_definition}
\end{figure}

\subsection{Multi-task Training for Timing Prediction} \label{sec:timing_training}

\begin{figure*}[t]
  \centering
  \includegraphics[width=\linewidth]{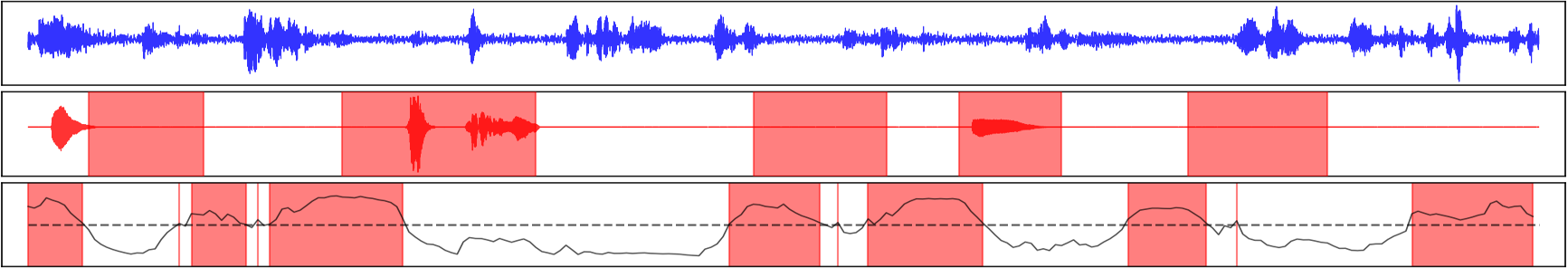}
  \Description{First row: speaker speech waveform; second: listener speech waveform with nodding intervals; third: nodding occurrence probability with predicted segments.}
  \caption{Example outputs of the timing prediction model. \emph{First}: speaker speech waveform. \emph{Second}: listener speech waveform with nodding intervals. \emph{Third}: probability of nodding occurrence with predicted segments.}\label{fg:timing_result}
\end{figure*}

The timing prediction module is trained on timing-related multi-task objectives. 
With the nodding timing prediction loss as the primary loss, training is performed via multi-task learning with the following five tasks.

\begin{itemize}
    \setlength{\itemsep}{0pt}
    \setlength{\parskip}{0pt}
    \item \textbf{Nodding Prediction}: Predicts the onset timing of listener nodding at each frame. As illustrated in Figure~\ref{fg:label_definition}, the ground-truth positive interval spans 500~ms before and after each nodding onset (1~s in total). The rationale is threefold: (1) onset timing suffices for the system to start nodding; (2) segment length varies with repetitions and speed, so each nodding event is treated equally; (3) a nod can be produced over an interval, not only at a single instant.
    \item \textbf{Voice Activity Projection (VAP)}: Predicts the presence or absence of speech in each of four future intervals within the next 2 seconds for each dialogue participant. The intervals are 0-200~ms, 200-600~ms, 600-1200~ms, and 1200-2000~ms, forming a $2~\text{channels} \times 4~\text{intervals}$ grid. The model outputs one of 256 classes representing the joint configuration.
    \item \textbf{Voice Activity Detection (VAD)}: Detects the presence or absence of speech for both participants at each frame. It complements the VAP predictions.
    \item \textbf{Backchannel Prediction (BP)}: Predicts verbal backchannel of the listener at each frame. Ground-truth labels shift verbal backchannel segments 500~ms earlier. Verbal backchannels often co-occur with nodding, so joint training should help nodding accuracy.
    \item \textbf{Backchannel Detection (BD)}: Detects the presence or absence of verbal backchannel of the listener at each frame. Like VAD for VAP, it complements BP by marking backchannel frames.
\end{itemize}

The loss function is defined as Equation~\eqref{eq:timing_mtl_loss}.
\begin{gather}
L = L_{nod} + w_{vap} L_{vap} + w_{vad} L_{vad} + w_{bp} L_{bp} + w_{bd} L_{bd} \label{eq:timing_mtl_loss}
\end{gather}
where $L_{nod}$, $L_{vap}$, $L_{vad}$, $L_{bp}$, and $L_{bd}$ are the frame-level cross-entropy losses for nodding prediction, voice activity projection, voice activity detection, backchannel prediction, and backchannel detection, respectively.
The weights $w_{vap}$, $w_{vad}$, $w_{bp}$, and $w_{bd}$ are hyperparameters for adjusting the weighting of loss terms, and all were set to 1 in our experiments.
To address the imbalance between positive and negative samples at the frame level for backchannel and nodding, the positive weight was set to 3 during training.
Training the timing prediction module also requires ground-truth verbal backchannel annotations in addition to nodding intervals.

\subsection{Fine-tuning for Kinematic Parameter Prediction} \label{sec:finetuning}

After training the timing prediction module, we fine-tune the kinematic parameter prediction module.
Training is performed via multi-task learning with the following four tasks, each corresponding to one kinematic parameter.
As illustrated in Figure~\ref{fg:label_definition}, the loss for each task is computed only over the positive intervals of the nodding timing prediction task.
Event-level kinematic parameters (Section~\ref{sec:data}) are assigned as labels to all frames in each positive interval.

\begin{itemize}
    \setlength{\itemsep}{0pt}
    \setlength{\parskip}{0pt}
    \item \textbf{Range Prediction}: Predicts the range of the head movement during nodding.
    \item \textbf{Speed Prediction}: Predicts the speed of the head movement during nodding.
    \item \textbf{Repetitions Prediction}: Predicts the number of nodding cycles. Formulated as a 3-class classification (1, 2, 3 or more).
    \item \textbf{Swing-up Prediction}: Predicts the presence or absence of swing-up. Formulated as a 2-class classification.
\end{itemize}

The loss function is defined as Equation~\eqref{eq:finetune_loss}.
\begin{gather}
L = w_{rp} L_{rp} + w_{sp} L_{sp} + w_{rep} L_{rep} + w_{sup} L_{sup} \label{eq:finetune_loss}
\end{gather}
where $L_{rp}$, $L_{sp}$, $L_{rep}$, and $L_{sup}$ represent the losses for range prediction, speed prediction, repetitions prediction, and swing-up prediction, respectively.
Range prediction and speed prediction use frame-level mean squared error (MSE), while repetitions prediction and swing-up prediction use frame-level cross-entropy.
Ground-truth range and speed are z-score normalized using the mean and standard deviation computed on the training data.
The weights $w_{rp}$, $w_{sp}$, $w_{rep}$, and $w_{sup}$ are hyperparameters, and all were set to 1 in our experiments.

For this fine-tuning step, the Self-attention and Cross-attention layers of the kinematic parameter prediction module are initialized with the weights learned by the timing prediction module and then updated on the objectives of Equation~\eqref{eq:finetune_loss}.
This improves kinematic parameter prediction by reusing timing features for turn-taking, verbal backchannel, and nodding onset.

\section{Experimental Evaluation} \label{sec:experiment}

\begin{table}[t]
  \centering
    \caption{Results for timing prediction}\label{tb:result_timing}
    \begin{tabular}{lccc} \hline
      Method&F1-score ($\uparrow$)&Precision ($\uparrow$)&Recall ($\uparrow$) \\ \hline
      Always&32.02$\pm$0.00&19.06$\pm$0.00&100.00$\pm$0.00 \\ \hdashline
      Stochastic&23.63$\pm$0.04&20.89$\pm$0.03&27.20$\pm$0.09 \\ \hdashline
      CPC&46.14$\pm$0.21&36.63$\pm$0.38&\phantom{1}62.34$\pm$1.43 \\ \hdashline
      Proposed&\textbf{52.19$\pm$0.40}&\textbf{41.93$\pm$0.60}&\phantom{1}\textbf{69.11$\pm$0.56} \\ \hline
    \end{tabular}
\end{table}

\begin{figure*}[t]
  \centering
  \includegraphics[width=\linewidth]{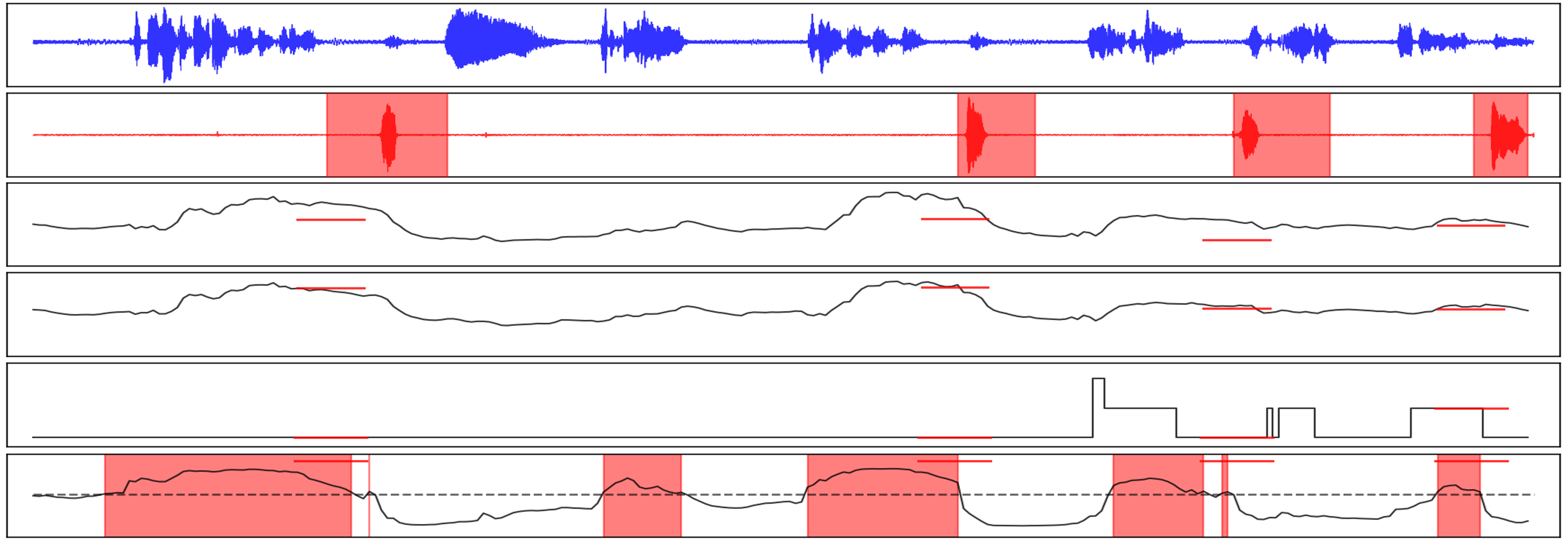}
  \Description{Six rows: speaker speech waveform; listener speech waveform with positive timing intervals shaded; predicted range, speed, repetitions, and swing-up. In rows~3--6, solid curves are model predictions and red lines are ground truth (constant within each positive interval).}
  \caption{Example outputs of the parameter prediction model. \emph{First row}: speaker speech waveform. \emph{Second row}: listener speech waveform with positive timing intervals shaded. \emph{Third--sixth rows}: predicted range, speed, repetitions, and swing-up. In rows~3--6, the solid curves show model predictions and the red lines show ground truth (constant within each positive interval).}\label{fg:param_result}
\end{figure*}

\begin{table*}[t]
  \centering
    \caption{Results for kinematic parameter prediction}\label{tb:result_param}
    \begin{tabular}{llcccccc} \hline
      \multicolumn{2}{c}{\multirow{2}{*}{Method}}&\multicolumn{2}{c}{Range (rad)}&\multicolumn{2}{c}{Speed (rad/s)}&Repetitions&Swing-up \\ \cline{3-8}
      &&MAE ($\downarrow$)&Corr.\ ($\uparrow$)&MAE ($\downarrow$)&Corr.\ ($\uparrow$)&Macro F1 ($\uparrow$)&F1-score ($\uparrow$) \\ \hline
      \multicolumn{2}{c}{Fixed}&0.0395$\pm$0.0000&0&0.0605$\pm$0.0000&0&22.25$\pm$0.00&0 \\ \hdashline
      \multicolumn{2}{c}{Stochastic}&0.0545$\pm$0.0003&0.0002$\pm$0.0000&0.0848$\pm$0.0005&0.0003$\pm$0.0000&33.30$\pm$0.01&24.10$\pm$0.11 \\ \hdashline
      \multirow{2}{*}{SVM}&F0+loud.&0.0377$\pm$0.0001&0.3314$\pm$0.0096&0.0571$\pm$0.0002&0.3565$\pm$0.0080&36.92$\pm$0.75&42.69$\pm$0.39 \\
      &Mimi&0.0367$\pm$0.0002&0.4083$\pm$0.0061&0.0559$\pm$0.0001&0.4103$\pm$0.0072&38.31$\pm$0.38&46.00$\pm$0.66 \\ \hdashline
      \multicolumn{2}{c}{CPC (Finetune)}&0.0370$\pm$0.0004&0.3774$\pm$0.0190&0.0568$\pm$0.0007&0.3851$\pm$0.0157&37.49$\pm$0.43&49.47$\pm$0.27 \\ \hdashline
      \multirow{2}{*}{Proposed}&Finetune&\textbf{0.0341$\pm$0.0003}&\textbf{0.5215$\pm$0.0127}&\textbf{0.0525$\pm$0.0003}&\textbf{0.5236$\pm$0.0111}&\textbf{39.37$\pm$0.89}&\textbf{53.45$\pm$0.72} \\
      &Scratch&0.0350$\pm$0.0004&0.4830$\pm$0.0224&0.0539$\pm$0.0008&0.4835$\pm$0.0245&37.22$\pm$1.01&50.79$\pm$0.34 \\ \hline
    \end{tabular}
\end{table*}

We evaluated the proposed model on two frame-level nodding tasks: timing prediction and kinematic parameter prediction.
For each task, the contribution of each auxiliary sub-task in multi-task training for timing prediction was examined through an ablation study.
The frame rate was 12.5~Hz, the context length 20~seconds, and we used five-fold cross-validation.

The Continuous Mimi Encoder used the weights of Mimi's Neural Encoder (ConvNet and Transformer layers), pretrained on approximately 7 million hours of speech data.
These parameters were frozen during training.
We also compared a CPC audio encoder, commonly adopted in the original VAP model~\cite{ekstedt2022} and its task-specific extensions~\cite{kato2025realtimenodding,inoue2025yeahunohcontinuous}, with weights pretrained on approximately 60,000 hours of LibriSpeech data and kept frozen during training.

\subsection{Timing Prediction} \label{sec:experiment_timing}

The first task is to predict whether nodding onset occurs at each frame.
The evaluation metrics were frame-level precision, recall and F1-score.

The following models were compared.
\begin{itemize}
    \setlength{\itemsep}{0pt}
    \setlength{\parskip}{0pt}
    \item (Always) Always predicting nodding in all frames.
    \item (Stochastic) Predicts nodding onsets by sampling the time between consecutive nod onsets from the interval distribution fit on the training data, with each nodding segment fixed to 1~second.
    \item (CPC) Using CPC as the audio encoder.
    \item (Proposed) Using the Continuous Mimi Encoder.
\end{itemize}

The results are presented in Table~\ref{tb:result_timing}.
The proposed model outperforms both Always, Stochastic and CPC in terms of F1-score.
This result indicates that the continuous features from the Mimi Audio Codec, prior to codebook quantization, are effective for predicting the onset timing of nodding.

Figure~\ref{fg:timing_result} shows output samples of the proposed model on the timing prediction task, demonstrating that it can effectively predict nodding onset earlier than the actual occurrence.

\subsection{Kinematic Parameter Prediction} \label{sec:experiment_param}

The next task is to predict the kinematic parameters of nodding.
The evaluation metrics are MAE and correlation coefficient (Corr.) for range and speed, macro F1-score for repetitions, and F1-score for swing-up.

In this task, we evaluated the proposed model and CPC, as well as support vector machine (SVM) baselines.
Support vector classification (SVC) was used for discrete tasks and support vector regression (SVR) for continuous-value tasks, both with the radial basis function (RBF) kernel, trained to predict the kinematic parameters of each nodding segment from the speech features of both speaker and listener over the preceding 2.5~seconds.

The following models were evaluated.

\begin{itemize}
    \setlength{\itemsep}{0pt}
    \setlength{\parskip}{0pt}
    \setlength{\parsep}{0pt}
    \item (Fixed) Predicting the fixed training-data mean for range and speed (0.06946~rad and 0.1046~rad/s); the fixed training-data modes for repetitions and swing-up (one repetition, no swing-up).
    \item (Stochastic) Sampling kinematic parameters from their joint training distribution.
    \item SVM
        \begin{itemize}
            \setlength{\itemsep}{0pt}
            \setlength{\topsep}{0pt}
            \setlength{\parsep}{0pt}
            \setlength{\parskip}{0pt}
            \setlength{\partopsep}{0pt}
            \item (F0+loud.) Using frame-level (100~Hz) F0 and loudness.
            \item (Mimi) Using Continuous Mimi Encoder features.
        \end{itemize}
    \item (CPC) Using CPC as the audio encoder while training the timing prediction model and fine-tuning for parameter prediction.
    \item Proposed
        \begin{itemize}
            \setlength{\itemsep}{0pt}
            \setlength{\topsep}{0pt}
            \setlength{\parsep}{0pt}
            \setlength{\parskip}{0pt}
            \setlength{\partopsep}{0pt}
            \item (Finetune) Initializing from the timing-trained module followed by fine-tuning for kinematic parameter prediction.
            \item (Scratch) Training the kinematic parameter prediction task from scratch.
        \end{itemize}
\end{itemize}

\begin{table*}[t]
  \centering
    \caption{Ablation study: scores for timing and kinematic parameter prediction tasks}\label{tb:ablation_scores}
    \begin{tabular}{lcccccccc} \hline
      \multirow{2}{*}{}&Timing&\multicolumn{2}{c}{Range (rad)}&\multicolumn{2}{c}{Speed (rad/s)}&Repetitions&Swing-up \\ \cline{2-8}
      &F1 ($\uparrow$)&MAE ($\downarrow$)&Corr.\ ($\uparrow$)&MAE ($\downarrow$)&Corr.\ ($\uparrow$)&Macro F1 ($\uparrow$)&F1 ($\uparrow$) \\ \hline
      Original&52.19$\pm$0.40&0.0341$\pm$0.0003&0.5215$\pm$0.0127&0.0525$\pm$0.0003&0.5236$\pm$0.0111&39.37$\pm$0.89&53.45$\pm$0.72 \\
      w/o VAD&52.16$\pm$0.21&0.0341$\pm$0.0004&0.5234$\pm$0.0160&0.0526$\pm$0.0003&0.5274$\pm$0.0097&39.19$\pm$0.70&53.28$\pm$0.46 \\
      w/o VAP&51.51$\pm$0.12&0.0345$\pm$0.0004&0.5034$\pm$0.0090&0.0529$\pm$0.0007&0.5106$\pm$0.0126&39.30$\pm$0.75&52.17$\pm$0.84 \\
      w/o BD&52.25$\pm$0.30&0.0339$\pm$0.0005&0.5278$\pm$0.0176&0.0526$\pm$0.0004&0.5229$\pm$0.0175&39.48$\pm$0.64&52.87$\pm$0.55 \\
      w/o BP&51.85$\pm$0.37&0.0338$\pm$0.0002&0.5275$\pm$0.0053&0.0525$\pm$0.0004&0.5228$\pm$0.0080&40.36$\pm$0.56&52.58$\pm$0.58 \\ \hline
    \end{tabular}
\end{table*}

The results are presented in Table~\ref{tb:result_param}.
Across all kinematic parameter prediction tasks, the proposed model outperforms Fixed, Stochastic, SVM, and CPC.
Furthermore, fine-tuning from the timing-trained module for kinematic parameter prediction substantially improved accuracy compared to training from scratch.
This indicates that features related to turn-taking, verbal backchannel, and nodding timing prediction are effective for predicting the form of nodding.
Compared with SVM with Mimi features, CPC scored lower on range, speed, and repetitions but higher on swing-up, suggesting that the proposed model's turn-taking structure contributes more to swing-up prediction accuracy.
% A possible reason is that the timing of cognitive shifts indicated by swing-up (e.g., reception of new information, surprise) is related to the turn-taking structure in dialogue.

Figure~\ref{fg:param_result} shows output samples of the proposed model on the parameter prediction task, demonstrating that it can appropriately predict kinematic parameters continuously.

\subsection{Ablation Study} \label{sec:ablation}

We also examined how much each auxiliary sub-task in multi-task timing training contributes to the two main tasks.
Table~\ref{tb:ablation_scores} presents the ablation results.
Removing the VAP loss lowered scores across all tasks from the Original model, suggesting that modeling turn-taking structure is critical for both timing and parameter prediction.
Relative to other parameters, swing-up prediction showed clearer F1 decreases when either the verbal backchannel prediction (BP) or detection (BD) loss was removed (from 53.45 to 52.58 and 52.87, respectively).
A possible reason for the sensitivity of swing-up to backchannel modeling is the close relationship between the timing of the listener's cognitive shift, indicated by swing-up (e.g., reception of new information, surprise)~\cite{mori2022cognitivenods}, and the timing of verbal backchannel.

\subsection{Real-time Processing Performance} \label{sec:exp:realtime}

The real-time processing performance of the proposed model was also evaluated.
Because the latency of the model is mostly caused by the Continuous Mimi Encoder, we created an ONNX (Open Neural Network Exchange) format (fp32) version of the Continuous Mimi Encoder for real-time processing.
Assuming deployment on laptop-class devices without GPUs, we measured the average inference time and the real-time factor (RTF) during inference on a CPU (Intel Core Ultra 9 275HX) over a 2-minute period.

\begin{table}[t]
  \centering
    \caption{Real-time processing performance}\label{tb:realtime}
    \begin{tabular}{lcc} \hline
      &Avg.\ inference time (s)&RTF \\ \hline
      w/o Mimi ONNX&0.05338&0.6673 \\
      w/ Mimi ONNX&0.04568&0.5711 \\ \hline
    \end{tabular}
\end{table}

The results are presented in Table~\ref{tb:realtime}.
The proposed model achieved an RTF of 0.67 without ONNX and 0.57 with ONNX.
Therefore, the proposed model is capable of real-time processing and, once integrated into avatar dialogue systems and conversational robots, can generate nodding in real time.

\section{Subjective Evaluation} \label{sec:subjective}

We integrated the proposed model into an avatar dialogue system with a CG agent\footnote{CG-CA Gene (c) 2023 by Nagoya Institute of Technology, Moonshot R\&D Goal 1 Avatar Symbiotic Society. \url{https://github.com/mmdagent-ex/gene}} (Figure~\ref{fg:system}).
In this system, diverse forms of nodding can be generated in real time in response to user utterances.
We conducted a subjective evaluation to investigate whether generating nodding with the proposed model gives a better impression to users compared to conventional methods.

\begin{table}[t]
  \centering
    \caption{Methods evaluated in subjective experiment}\label{tb:subjective_methods}
    \begin{tabular}{clcc} \hline
      &Method&Timing&Parameters \\ \hline
      A)&No Nodding&--&-- \\
      B)&Stochastic + Fixed&Stochastic&Fixed \\
      C)&Proposed + Fixed&Proposed&Fixed \\
      D)&Proposed + Stochastic&Proposed&Stochastic \\
      E)&Proposed + Proposed&Proposed&Proposed \\ \hline
    \end{tabular}
\end{table}

\begin{figure}[t]
  \centering
  \includegraphics[width=\linewidth]{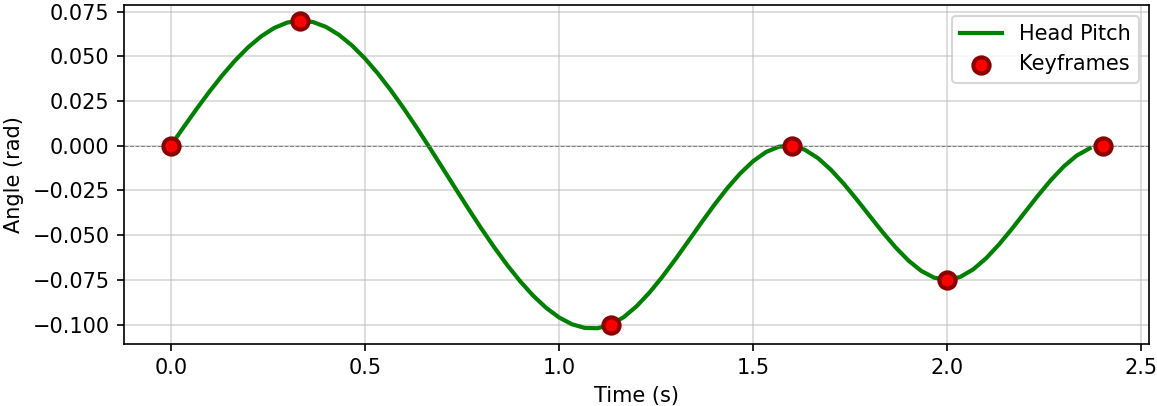}
  \Description{Keyframes and interpolated nodding head motion generated from predicted parameters.}
  \caption{Nodding motion generation from predicted kinematic parameters via cubic spline interpolation}\label{fg:motion_gen}
\end{figure}

\subsection{Method}

\begin{figure*}[t]
  \centering
  \includegraphics[width=0.95\linewidth]{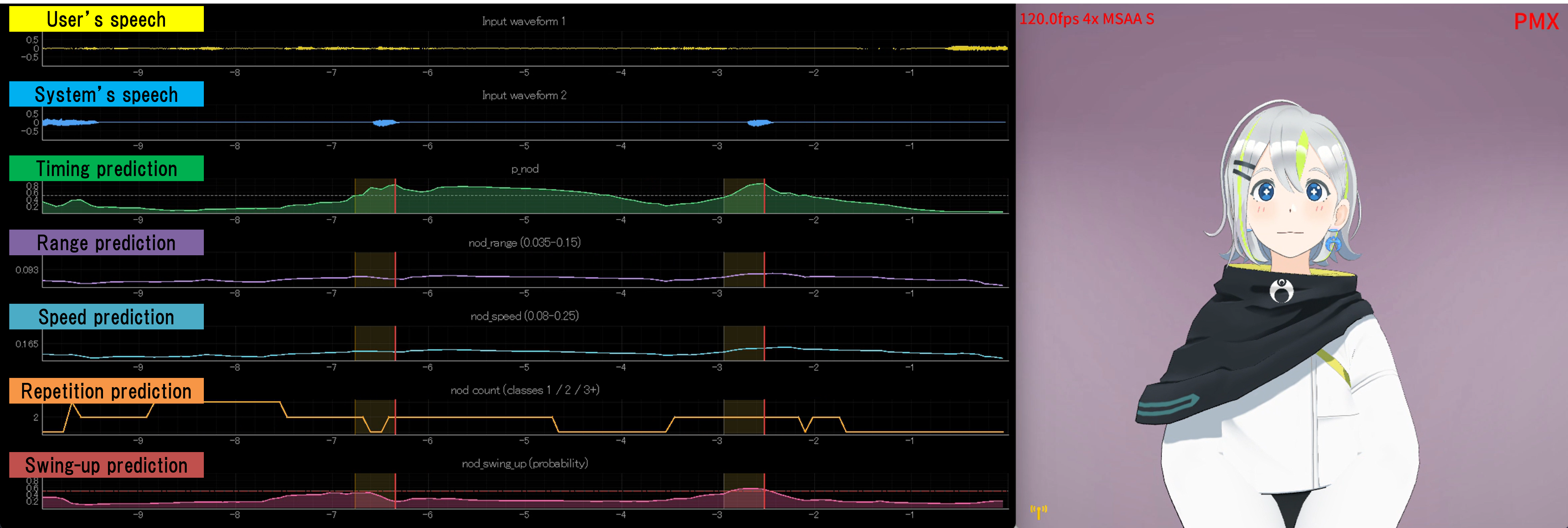}
  \Description{Screenshot of the avatar dialogue system with a CG agent integrated with the proposed model.}
  \caption{Avatar dialogue system integrated with the proposed model}\label{fg:system}
\end{figure*}

We evaluated the five methods in Table~\ref{tb:subjective_methods}; \textbf{Stochastic} and \textbf{Fixed} follow the same timing and parameter baselines as in Section~\ref{sec:experiment}.
% In Method~B, nod timing was sampled from the interval distribution fit on the training data at each run.
% In Method~D, each kinematic parameter was sampled independently from its training distribution whenever a nod was predicted.
% Methods~B and C used fixed nod parameters, with training-data means for range and speed (0.06946~rad and 0.1046~rad/s) and modes for discrete attributes (one repetition, no swing-up).
In Methods~D and E, at runtime (Figure~\ref{fg:system}), when the nodding probability exceeds a threshold continuously for a fixed number of frames (e.g., $K{=}5$ frames at 12.5~Hz, equivalent to 400~ms), we average the parameter predictions over those frames to obtain the event-level parameter set.
A full nodding trajectory satisfying these parameters is then synthesized by rule-based cubic-spline interpolation (Figure~\ref{fg:motion_gen}) and sent to the avatar.
Because the positive interval is defined as $\pm$500~ms around each onset, the model predicts nods sufficiently earlier than their actual occurrence (Figure~\ref{fg:timing_result}), so requiring a short run of positive frames yields stable detection without perceptible delay.

We recorded 45 avatar videos by passing nine sessions randomly chosen from held-out test sessions (five-fold CV, Section~\ref{sec:experiment}) through each of five variants (9 sessions $\times$ 5 methods).

The nine sessions were divided into 3 groups, and 30 crowdsourcing workers were recruited per group, totaling 90 participants.
Each participant watched 15 videos (3 sessions $\times$ 5 methods) plus one dummy video for attention check (16 total).
For this attention check, the last 5 seconds of the dummy video instructed participants to rate all items as 1, and we excluded participants who did not follow the instruction.
Each participant evaluated each video on the following 7 metrics.

\begin{enumerate}
    \setlength{\itemsep}{0pt}
    \setlength{\parskip}{0pt}
    \item How human-like is the avatar's response? (\textbf{human likeness})
    \item How natural is the avatar's response? (\textbf{naturalness})
    \item How varied (non-monotonous) is the response? (\textbf{diversity})
    \item How well does the avatar appear to be listening to the user? (\textbf{attentiveness})
    \item How well does the avatar encourage continued speaking? (\textbf{facilitation})
    \item How well does the avatar appear to understand? \\(\textbf{understanding})
    \item How well does the avatar appear to empathize? (\textbf{empathy})
\end{enumerate}
Each metric was scored on a 7-point scale, videos were shown in random order, and the experiment was conducted in Japanese.

\subsection{Results and Analysis}

Each participant evaluated three videos per method, and the per-method score for each metric was computed by averaging these three ratings.
After the attention check, \textit{n}=60~participants remained.
Tables~\ref{tb:subjective_scores} and~\ref{tb:subjective_pvalues} present the mean scores and Bonferroni-corrected post-hoc pairwise comparisons, respectively.
A mixed-design ANOVA with group (3 levels) as a between-subjects factor and method (5 levels) as a within-subjects factor revealed a significant main effect of the method for all seven metrics (Greenhouse-Geisser corrected $p < .001$; partial $\eta^2 \approx .78$-.85).
The main effect of the group was not significant for any metric ($p > .14$).
The group $\times$ method interaction was significant for naturalness ($p = .002$), attentiveness ($p = .042$), and understanding ($p = .043$).

% \begin{figure}[t]
%   \centering
%   \includegraphics[width=\linewidth]{figure/compare_range_speed.png}
%   \Description{Range versus speed for Methods D and E and ground truth, with dashed regression lines.}
%   \caption{Range versus speed for nodding under Method~D, Method~E (proposed), and ground truth (dashed linear regression).}\label{fg:comparison_range_speed}
% \end{figure}

{\tabcolsep = 0.12cm
\begin{table}[t]
  \centering
    \caption{Averaged scores of subjective evaluation (\textit{n}=60). Hum: human likeness; Nat: naturalness; Div: diversity; Att: attentiveness; Fac: facilitation; Und: understanding; Emp: empathy. Method names match Table~\ref{tb:subjective_methods}.}\label{tb:subjective_scores}
    {\small
    \begin{tabular}{clccccccc} \hline
      \multicolumn{2}{c}{} & \multicolumn{3}{c}{Response} & \multicolumn{4}{c}{Avatar} \\
      \cline{3-5}\cline{6-9}
      & Method & Hum & Nat & Div & Att & Fac & Und & Emp \\ \hline
      A) & No nodding & (1.64) & (1.59) & (1.09) & 1.94 & 1.13 & 1.87 & 1.41 \\
      B) & Stochastic $+$ Fixed & 4.19 & 4.06 & 3.11 & 4.34 & 3.73 & 4.23 & 4.08 \\
      C) & Proposed $+$ Fixed & 4.53 & 4.52 & 2.94 & 4.98 & 4.26 & 4.72 & 4.51 \\
      D) & Proposed $+$ Stochastic & \underline{5.38} & \underline{5.24} & \textbf{5.08} & \underline{5.66} & \textbf{5.05} & \underline{5.43} & \underline{5.44} \\
      E) & Proposed $+$ Proposed & \textbf{5.46} & \textbf{5.42} & \underline{4.87} & \textbf{5.69} & \underline{5.02} & \textbf{5.45} & \textbf{5.49} \\ \hline
    \end{tabular}
    }
\end{table}
}

{\tabcolsep = 0.08cm
\begin{table}[t]
  \centering
    \caption{Post-hoc pairwise comparison \textit{p}-values (Bonferroni corrected; \textit{n}=60). Column abbreviations as in Table~\ref{tb:subjective_scores}.}\label{tb:subjective_pvalues}
    {\small
    \begin{tabular}{cccccccc} \hline
      Comparison&Hum&Nat&Div&Att&Fac&Und&Emp \\ \hline
      B vs.\ C&\phantom{<}.030*\phantom{*}&\phantom{<}.023*\phantom{*}&\phantom{<}1.00\phantom{**}&<.001**&\phantom{<}.002**&\phantom{<}.002**&\phantom{<}.002** \\
      C vs.\ E&<.001**&<.001**&<.001**&<.001**&<.001**&<.001**&<.001** \\
      D vs.\ E&\phantom{<}1.00\phantom{**}&\phantom{<}.205\phantom{**}&\phantom{<}.742\phantom{**}&\phantom{<}1.00\phantom{**}&\phantom{<}1.00\phantom{**}&\phantom{<}1.00\phantom{**}&\phantom{<}1.00\phantom{**} \\ \hline
      \multicolumn{8}{r}{*p<0.05\quad **p<0.01} \\
    \end{tabular}
    }
\end{table}
}

\begin{figure*}[t]
  \centering
  \includegraphics[width=\linewidth]{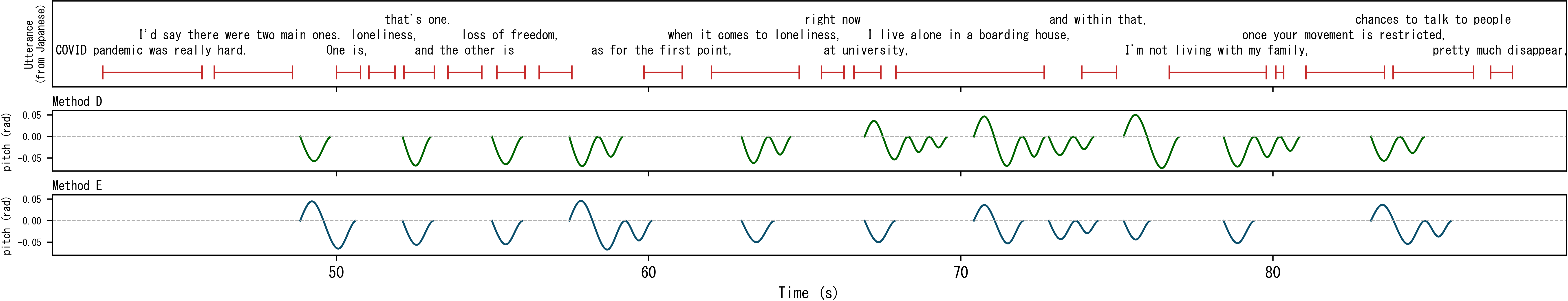}
  \Description{Comparison of Methods D and E: utterance timeline and head pitch angle over time for generated nodding.}
  \caption{Comparison of nodding motions generated by Method~D (stochastic parameters) and Method~E (proposed): utterance timeline and head pitch (rad) over time}\label{fg:qualitative}
\end{figure*}

Across all metrics, Methods~D and E, which predict nodding timing with the proposed model and generate diverse forms of nodding, tended to achieve the highest scores.
Comparing B and C, proposed timing significantly outperformed stochastic timing on all metrics except diversity.
In particular, a large difference was observed in attentiveness, indicating that proposed nodding-onset timing prediction enhances the attentiveness of dialogue systems.
Comparing C and E, generating diverse forms of nodding yielded significantly higher scores than generating fixed nodding on all metrics.
This indicates that a conversational avatar expressing diverse forms of nodding according to context can behave more naturally and human-like, and can better express understanding and empathy.
Comparing D and E, diversity was slightly higher for Method~D and naturalness slightly higher for Method~E descriptively, but no significant differences were found on any metric.
These results demonstrate that, by predicting both timing and kinematic parameters of nodding in real time with the proposed model and expressing diverse forms of nodding at appropriate timing, it is possible to enhance the naturalness and human-likeness of the responses and to convey attentive listening, understanding, and empathy.

\subsection{Qualitative Analysis}

We conducted a qualitative analysis of nodding behavior under Methods~D and E, for which the subjective evaluation found no significant differences on any metric.
In the example in Figure~\ref{fg:qualitative}, the nodding sequences generated by the proposed method (Method~E) exhibit the following tendencies:

\begin{itemize}
    \setlength{\itemsep}{0pt}
    \setlength{\parskip}{0pt}
    \item When the speaker's talk is likely to continue, a single small nod is produced as a continuer.
    \item When the talk appears to be ending, or when the user presents new information, swing-up tends to accompany the nod, and the range is larger.
\end{itemize}

Specifically, in Method~E, while the speaker describes two main difficulties during the pandemic (about 52-55~s), small single nods continued as continuers.
After both ``loneliness'' and the second difficulty ``loss of freedom'' had been introduced, a relatively large two-cycle nod with swing-up occurred around 57~s in response to that new information.
During the subsequent explanation of loneliness (about 63-68~s), small single nods continued, and around 70~s a swing-up nod appeared when the speaker disclosed living alone.
Around 82~s, a swing-up nod followed talk of loneliness causes (e.g., ``I'm not living with my family'', ``once your movement is restricted.'').
This can be viewed as partial uptake or alignment while the speaker explains why loneliness arises.
These patterns align with prior analyses that listener swing-up nods often accompany understanding of the speaker's intent, receipt of new information, or a shift in cognitive state~\cite{mori2022cognitivenods, navarretta2012feedback, whitehead2011some}.
By contrast, under Method~D, large swing-up nods occurred around 67 and 75~s at moments when the topic still seemed likely to continue, which appears less appropriate for a listener in context.

Such characteristics of the proposed model were not clearly reflected in the subjective evaluation results.
This is likely because participants watched pre-recorded videos as third-party observers rather than interacting with the avatar directly.
In such a format, subtle qualities may be difficult to assess, in particular whether the avatar appears to understand the speaker and respond with empathy.
Additionally, since five methods were evaluated simultaneously, the overall evaluation may have been dominated by factors that differ substantially across methods, such as how diverse nodding motions are in each condition and whether timing is appropriate or not.
To more clearly demonstrate that the proposed model produces nodding forms that match the conversational content, further experiments such as paired comparisons in face-to-face interactions are needed.

\section{Conclusion} \label{sec:conclusion}

In this paper, we have proposed a model that predicts in real time both the timing and kinematic parameters of nodding, an important nonverbal listener response, from the speech signals of both the speaker and the listener.
Each of the timing and parameter modules employs a dyadic attention network over the speaker and listener channels.
We fine-tune the kinematic parameter prediction module from the trained timing prediction module.
As a result, we achieved an F1 score of 52.19 on timing prediction and, on kinematic parameter prediction, correlations of 0.5215 and 0.5236 for range and speed, macro F1 39.37 for repetitions, and F1 53.45 for swing-up.
Furthermore, the effectiveness of this fine-tuning procedure was demonstrated.
In addition, the ablation study showed that multi-task learning with the VAP task is important for timing prediction, and that auxiliary tasks related to verbal backchannels also influence swing-up prediction accuracy.
We showed that the proposed model is capable of real-time processing and integrated it into avatar dialogue systems and conversational robots.
Subjective evaluation experiments demonstrated that the proposed method significantly outperformed both the stochastic timing baseline and the fixed-motion nodding baseline.

Future work includes advancing from predicting nodding kinematic parameters to generating natural nodding motions in a context-aware manner from kinematic parameters, and building models that predict listener nonverbal responses using not only speech but also the speaker's visual nonverbal cues (e.g., eye gaze, nodding) during dialogue.

\section*{Safe and Responsible Innovation Statement}
Our models estimate listener nodding from conversational speech to drive attentive-listening interfaces.
We expect a relatively small misuse footprint compared with unconstrained generative systems, but interfaces must not suggest that a live human steers nodding when it is model-generated, including claims that an avatar or robot is entirely human-operated.
Speech materials and derived features are handled under local access controls.
The dialogue corpus was collected with informed consent for research use and sharing, and released recordings exclude directly identifying items.

\begin{acks}
This work was supported by JST Moonshot R\&D JPMJPS2011 and JST PRESTO JPMJPR24I4.
K.~Kato was a scholarship recipient of the Iwadare Scholarship Foundation in fiscal year 2025.
\end{acks}

{\footnotesize
\bibliographystyle{unsrt}
\bibliography{sample-base}
}

\end{document}